\documentclass[aps, 12pt]{revtex4}

\begin{document}
\title{Stieltjes Electrostatic Model Interpretation for Bound State Problems}
\author{K. V. S. Shiv Chaitanya}
\affiliation{BITS Pilani, Hyderabad Campus, Jawahar Nagar, Shameerpet Mandal,
Hyderabad, India.}

\begin{abstract}
In this paper, Stieltjes electrostatic model and quantum Hamilton Jacobi formalism is analogous to each other is shown. This analogy allows, the bound state problem to mimics as $n$ unit moving  imaginary charges $i\hbar$, which are placed in between the two fixed imaginary charges arising due to the classical turning points of the potential. The interaction potential between $n$ unit moving imaginary charges $i\hbar$ is given by logarithm of the wave function. For an exactly solvable potential, this system attains stable equilibrium position at the zeros of the orthogonal polynomials depending upon the interval of the classical turning points.
\end{abstract}
\maketitle
\textbf{keywords :} Orthogonal polynomials, quantum Hamilton Jacobi and zeros of orthogonal polynomials.

\section{Introduction}

Stieltjes \cite{st,st1} considered the following problem with $n$ moving unit charges, interacting  through a logarithmic potential, are placed
between  two fixed charges $p$ and $q$ at $-1$ and $1$ respectively on a real line.
He then proved that the system attains a stable equilibrium position at the zeros of the Jacobi polynomial $P^{(\alpha,\beta)}_n(x)$. Proof is given in Szego's book (section 6.7) \cite{sz}. If, the interval is changed on the real line, for the fixed charges, then the the system attains stable equilibrium position at the zeros of the orthogonal polynomial with the respective intervals. For example, in the interval $[0;\infty)$ one gets the Laguerre polynomials $L^{(k)}_n(x)$ and for the the interval $(-\infty;\infty)$ one gets the Hermite polynomials polynomials $H_n(x)$. This model has been extended to the zeros of general orthogonal polynomials in the ref \cite{mhd}.

The Quantum Hamilton Jacobi (QHJ) formalism, was formulated for the bound state problems by Leacock and Padgett \cite{qhj, qhj1} and later on was successfully applied to several exactly
solvable models (ESM) \cite{bhal, bhal1, sree, sree1, geo} in one dimension, the quasi - exactly solvable (QES) models \cite{geo1},  the periodic potentials \cite{sree2} and the PT symmetric potentials \cite{sree3} in quantum mechanics. In QHJ 
the central role is played by the quantum momentum function (QMF). This function, in general, contains fixed poles that arises due to the classical turning points of the potential. In general, for most of the potentials in quantum mechanics there will be only two fixed poles, and $n$ moving poles arise due the zeroes of wave function. Thus, one can immediately see the connection between the two scenarios presented above. The fixed poles of the potential are like the the two fixed charges and $n$ moving poles on the real line are like $n$ moving charges. 

\subsection{Electrostatic Model}
Stieltjes considered the interaction forces  for the $n$ moving unit charges arising from a logarithmic potential which are in between the to fixed charges $p$ and $q$ at $-1$ and $1$ respectively on a real line as

\begin{eqnarray}
L&=&- Log D_n(x_1, x_2...x_n)+p\sum_{i=1}^n Log(\frac{1}{\vert 1-x_i\vert})\nonumber\\&&+q\sum_{i=1}^n Log(\frac{1}{\vert 1+x_i\vert}),\label{st}
\end{eqnarray}
where
\begin{eqnarray}
- Log D_n(x_1, x_2...x_n)=\sum_{1\leq i <j\leq n}^n Log(\frac{1}{\vert x_i-x_j\vert})
\end{eqnarray}

Then, he proved in ref \cite{st, st1} that the expression (\ref{st}) becomes a minimum 
\begin{eqnarray}
\sum_{i=1,i\neq k}^n \frac{1}{x_i-x_k}-\frac{p}{x_k-1}-\frac{q}{x_k+1}=0.\label{es1}
\end{eqnarray}
when $(x_1, x_2, \cdots,x_n)$ are the zeros of the Jacobi polynomial
\begin{eqnarray}
(1-x^2)P_n''(x)+2[q - p - (p + q)x]P_n'(x)=n[n+ 2(p + q) - 1]P_n(x),
\end{eqnarray} 
where $P^{(2p-1,2q-1)}_n(x)$ are the Jacobi polynomial. For the proof refer to Szego's book (section 6.7) \cite{sz}.
The zeros of the Laguerre and the Hermite polynomials admit the same interpretation.

\subsection{Quantum Hamilton Jacobi}

In this section, a brief review of  Quantum Hamilton Jacobi formalism is presented below.
For details see the references \cite{geo, sree}. 
The Schr\"odinger equation is given by,
\begin{equation}
- \frac{\hbar^2}{2m}\nabla^2\psi(x,y,z)+ V(x,y,z) \psi(x,y,z) = E
  \psi(x,y,z).    \label{sc} 
\end{equation}
One defines a function $S$ analogous to the classical characteristic function by the relation 
\begin{equation}
\psi(x,y,z) = \exp\left(\frac{iS}{\hbar}\right)       \label{ac}
\end{equation}
which, when substituted in (\ref{sc}), gives
\begin{equation}
(\vec{\nabla}S)^2 -i \hbar \vec{\nabla}.(\vec{\nabla}S) = 2m (E
  - V(x,y,z)).   \label{qhj0} 
\end{equation}
the quantum momentum function $p$ is defined in terms of the function $S$ as
\begin{equation}
\vec{p} = \vec{\nabla} S. \label{mp}
\end{equation}
Substituting (\ref{mp}) in (\ref{qhj0}) gives the QHJ equation
for $\vec{p} $ as 
\begin{equation}
(\vec{p})^2 - i \hbar \vec{\nabla}.\vec{p} = 2m (E - V(x,y,z))
  \label{bhy} 
\end{equation}
and from (\ref{sc}) and (\ref{mp}), one can see that $\vec{p}$ 
is the the logarithmic derivative of $\psi(x,y,z)$ {\it i. e}. 
\begin{equation}
\vec{p} = -i \hbar \vec{\nabla} ln \psi(x,y,z) \label{lg}
\end{equation}
The above discussion of the QHJ formalism is done in three
dimensions the same equation in one dimension takes the following form 
\begin{equation}
p^2 - i \hbar \frac{dp}{dx} = 2m (E - V(x)),       \label{qhj1}
\end{equation}
which is also known as the Riccati equation. In one dimension the eq (\ref{lg}) take the form
\begin{equation}
p = -i \hbar \frac{d}{dx}ln \psi(x). \label{lg1}
\end{equation}
It is shown by Leacock and Padgett \cite{qhj, qhj1} that the action angle variable 
gives rise to exact quantization condition
\begin{equation}
J(E) \equiv  \frac{1}{2\pi} \oint_C{pdx} = n\hbar.       \label{act}
\end{equation}
\section{Model }

By considering the form of the wave function, in the equation (\ref{lg1}), to be  $\psi=\prod_{i=1}^N(x-x_i)$. Then, in the quantum momentum function this corresponds to $n$ zeros on the real line. These zeros are also called the moving poles in the language of QHJ. Choosing, an exactly solvable potential $V(x)$, with two fixed poles as the classical turning points, substituting in equation (\ref{qhj1}). Then, for bound states 
the following feature always arises in QHJ that the $n$ moving poles lie in between the two fixed poles and the solutions are the orthogonal polynomials for the  exactly solvable potential $V(x)$. The examples are the Harmonic oscillator, the Coulomb potential , the Scarf potential etc \cite{sree, geo}.

Thus, the connection between the QHJ and the Stieltjes electro static model can be seen. The fixed poles of the potential are like the two fixed charges and the $n$ moving poles of the real line are like $n$ moving charges. In the electrostatic model the moving charges interact with the logarithmic potential and in the QHJ  the logarithmic potential arises from the wave function. As the quantum momentum function is log derivative of the wave function.

Starting with the QMF the analogue between the two models is established.  The fact that only the residues of the QMF are required for
finding the eigenvalues is studied in ref \cite{bhal,bhal1}. The formalism for effectively obtaining both the eigenfunctions and the eigenvalues from the singularity structure of the quantum momentum function is given in ref \cite{sree}. The quantum momentum function assume that there are no other singular points of $p$ in
the complex plane.
Then the quantum momentum function is given by \cite{bhal, bhal1, sree, sree1, geo, geo1}
\begin{equation}
p=\sum_{k=1}^n\frac{-i}{x-x_k}+Q(x),\label{uf}
\end{equation}
here the moving poles are simple poles with residue $-i\hbar$ (we take here $\hbar=m=1$) \cite{sree, geo} and $Q(x)$ is the residues of fixed poles arising due to the exactly solvable potentials. 
This equation resembles the equation (\ref{es1}) except that it is the minimum of the potential.
Thus, the  quantum momentum function can interpret as system of equations arising for the logarithmic derivative of wave function and fixed poles arising from the classical turning points. By asking the following question, when does this system come to stable equilibrium ?
From the above discussion it is clear that answer can be obtained using Stieltjes Electrostatic model. It can be shown that the same wave function can be obtained from both the models. Thus, their exist a analogy  between the Stieltjes electrostatic interpretation for zeros of orthogonal polynomials and the quantum Hamilton Jacobi formalism.

The most important point in the quantum Hamilton Jacobi formalism is that if one has the total information 
about the pole structure of the quantum momentum function than by calculating the integral in the eq (\ref{act}) one gets the exact quantization condition. Or one can also get the quantization condition by converting the quantum momentum function  into a differential equation.  Therefore, the connection between the Stieltjes electrostatic interpretation and the quantum Hamilton Jacobi formalism is established by solving the quantum momentum function as a differential equation. This is achieved  by solving for $lim_{x\rightarrow x_k}ip(x)=0$ and thus the
 equation (\ref{uf}) is given by
\begin{equation}
\lim_{x\rightarrow x_k}\left[\sum_{k=1}^n\frac{1}{x-x_k}+iQ(x)\right]=0.\label{uf11}
\end{equation}
By introducing the polynomial
\begin{eqnarray}
f_n(x)=(x-x_1)(x-x_2)\cdots (x-x_n),
\end{eqnarray}
then using the following relation \cite{mhd, mhd1},
\begin{eqnarray}
\sum_{j=1,i\neq k}^n \frac{1}{x_j-x_k}&=&\lim_{x\rightarrow x_k}
\left[\frac{f_n'(x_k)}{f_n(x_k)}-\frac{1}{x-x_k}\right].\label{lhp}
\end{eqnarray}
As $Q(x)$ does not have any poles at $x_k$, the equation (\ref{uf11}) is given as
\begin{equation}
\sum_{j=1,i\neq k}^n \frac{1}{x_j-x_k} +iQ(x)=0.\label{uf1}
\end{equation}
It is clear that above equation is similar to eq (\ref{es1}). Therefore, the
Stieltjes electrostatic method goes through for solving the quantum momentum function.
By using the formula \cite{mhd, mhd1} 
\begin{eqnarray}
2\sum_{j=1,i\neq k}^n \frac{1}{x_k-x_j}&=&\frac{f_n''(x_k)}{f_n'(x_k)},
\end{eqnarray}
then the equation (\ref{uf1}) becomes 
\begin{eqnarray}
-\frac{1}{2}\frac{f_n''(x_k)}{f_n'(x_k)}+iQ(x_k)=0.\label{dif} ~~~~1<k<n
\end{eqnarray} 
By demanding the solution equation (\ref{dif}), for an exactly solvable potentials, to be zeros of certain orthogonal polynomials makes the points $x_k$ to vanish. The interval is fixed by the fixed poles of the potential. It is well known that the classical orthogonal polynomials arise as solutions to the bound states problems. Thus, the classical orthogonal polynomials are classified into three different categories depending upon the range of the polynomials. The polynomials in the intervals $(-\infty;\infty)$ are the Hermite polynomials, in the intervals $[0;\infty)$ are the Laguerre polynomials and in the intervals $[-1;1]$ are the Jacobi polynomials. Their singularity structure is as follows $Q(x)=x$,  $Q(x)=\frac{b}{x}+C$, and $Q(x)=-\frac{a}{x-1}-\frac{b}{x+1}$ for the Hermite, the Laguerre and the Jacobi polynomials respectively. Hence, the differential equation can be obtained by examining at the singularity structure of the quantum momentum function. This can be seen by rewriting the
equation (\ref{dif}) as 
\begin{eqnarray}
-f_n''(x) + 2iQ(x_k)f_n'(x)=0.\label{dif1} ~~~~1<k<n
\end{eqnarray} 
The function $Q(x)$ which has the information of fixed pole singularity structure appears as the coefficient of $f_n'(x_k)$. By examining the differential equations of the Hermite, the Laguerre and the Jacobi polynomials the coefficients of $Q(x)$ are fixed.

Let $f(x)=L_{\lambda}^{m} (x)$ satisfy the Laguerre differential equation 
\begin{equation}
  x\frac{d^2}{dx^2}f (x) + (m+1-x) \frac{d}{dx}f (x) + \lambda  f (x) = 0,\label{lague}
\end{equation}
where $\lambda$ is an integer. By examining the first two terms of the differential equations (\ref{dif1}) and (\ref{lague}) one gets
\begin{eqnarray}
2iQ(x)=\frac{(m+1)}{x}-1
\end{eqnarray}
the singularity structure for the Laguerre is
 \begin{eqnarray}
Q(x)=\frac{b}{x}+C
\end{eqnarray}
thus one gets $b=-i(m+1)$ and $C=i$. Similarly for the Jacobi differential equation 
\begin{eqnarray}
(1-x^2)f_n''(x_k)+2[p - q - (p + q)x]f_n'(x_k)+n[n+ 2(p + q) - 1]f_n(x)=0
\end{eqnarray} 
again comparing the first two terms
\begin{eqnarray}
2iQ(x)=-\frac{p}{x_k-1}-\frac{q}{x_k+1}
\end{eqnarray}
the singularity structure for the Jacobi is
 \begin{eqnarray}
2iQ(x)=-\frac{a}{x-1}-\frac{b}{x+1}
\end{eqnarray}
thus one has $p=-ia$ and $q=-ib$.
Similar analysis can be done for the Hermite polynomials.

The values of $m$, $p$ and $q$ has to be determined as these are not points
like in the electrostatic model. The method adopted by QHJ, search for the polynomial solutions leads to quantization, are used to calculate these values.
By writing the quantum  momentum function as
\begin{equation}
p=\sum_{k=1}^n i\frac{f'(x)}{f(x)}+Q(x)\label{uf4}
\end{equation}
and substituting in (\ref{qhj1}) then one gets
\begin{eqnarray}
f_n''(x_k) + 2iQ(x)f_n'(x_k)+[Q^2(x)-iQ'(x)-E+V(x)]f(x)=0.\label{dif11}
\end{eqnarray} 
The first two terms in the above differential equation arises due to the pole structure. Now by fixing the solution to be certain orthogonal polynomial 
depending upon the pole structure of $Q(x)$. This is equivalent to demanding $ [Q^2(x)-iQ'(x)-E + V(x)]$ to be constant i.e. "the search for the polynomial solutions leads to quantization". This will fix the values of the residues appearing for fixed poles and in the process the system is quantized for a given $V(x)$. Thus by comparing the equation (\ref{dif1}) and (\ref{dif11}) it can seen that the singularity structure of $iQ(x)$ determines the differential equation. Therefore, the same wave function is obtained from both the methods.

\section{Discussion}
From the previous discussion, it is clear that Stieltjes electrostatic model and quantum Hamilton Jacobi formalism are analogous to each other. Therefore,
this analogy allows,  the bound state problem to mimics as $n$ unit moving  imaginary charges $i\hbar$, which are placed in between the two fixed imaginary charges arising due to the classical turning points of the potential. The interaction potential between $n$ unit moving imaginary charges $i\hbar$ is given by logarithm of the wave function. For an exactly solvable potential, this system attains stable equilibrium position at the zeros of the orthogonal polynomials depending upon the interval of the classical turning points. Once charges arise  in any model they satisfy the continuity equation of the form
\begin{eqnarray}
\frac{\partial}{\partial t}\rho +\nabla \cdot J=0.
\end{eqnarray}
 Since, the equation (\ref{dif1}) and (\ref{dif11}) are nothing but the different form of the Schroedinger equation. Therefore their exist a continuity equation of this form for these imaginary with $\rho=\int_V\psi^*\psi dV$ is probability density function and $J=\frac{\hbar}{i}[\psi^*(\nabla\psi)-\psi(\nabla\psi^*)]$ is probability current density function. 
Hence, the conservation of probability leading to conservation of imaginary charge and probability current leads to current density for imaginary charge.
In this model $\rho$ is the amount of imaginary charge and $J$ is the current density for imaginary charge. Thus, this model is consistent with quantum mechanics.

\section{Conclusion}

In this paper, the two different models, one the Stieltjes electrostatic model and the other one Quantum Hamilton Jacobi formalism are examined. Except that one is a classical model and another is a quantum model. It is shown that Stieltjes electrostatic model and  quantum Hamilton Jacobi formalism are analogous to each other. One new feature comes out of this study is that the wave function can be obtained from the quantum momentum function itself, one  need not solve the quantum Hamilton Jacobi equation. From Stieltjes electrostatic model gives nice insights to the methodology of quantum Hamilton Jacobi formalism. It is interesting to note that the Stieltjes electrostatic model existed almost 30 years before quantum mechanics came into existence.
\section*{Acknowledgments}
Author thank  A. K. Kapoor, V. Srinivasan, Prasanta K. Panigrahi, Sashideep Gutti and P K Thirivikraman for stimulating conversations. 

\end{document}